\documentclass[aps,amssymb,amsmath,pra,twocolumn,showpacs,superscriptaddress,longbibliography]{revtex4-2}

\usepackage{graphicx}
\usepackage{dcolumn}
\usepackage{bm}
\usepackage{subfigure}
\usepackage{color}

\usepackage{amssymb,amsmath,amsfonts,latexsym,graphicx,verbatim}

\usepackage[margin=0.67in]{geometry}

\usepackage[english]{babel}
\usepackage{times} 
\usepackage{dsfont}
\usepackage{latexsym}
\usepackage{fancyhdr}
\usepackage{float}
\usepackage{afterpage}
\usepackage{enumitem}
\usepackage{mleftright}

\usepackage{eso-pic, graphicx}

\usepackage{listings}
\usepackage{multirow}
\usepackage{xcolor,colortbl}

\usepackage{bbm}
\usepackage{upgreek}
\usepackage{amsmath}
\usepackage{newtxtext,newtxmath}

\definecolor{Ablue}{rgb}{0.96,0.24,0.00}

\definecolor{Abluetitle}{rgb}{0.,0.24,0.51}

\definecolor{orange}{rgb}{0.96,0.24,0.00}

\definecolor{darkred}{rgb}{0.55, 0.0, 0.0}

\definecolor{darksalmon}{rgb}{0.91, 0.59, 0.48}
\definecolor{maroon}{cmyk}{0,0.87,0.68,0.32}

\definecolor{mustard}{rgb}{1.0, 0.86, 0.35}

\definecolor{Gray}{gray}{0.85}
\definecolor{LightCyan}{rgb}{0.88,1,1}
\newcolumntype{a}{$$>$${\columncolor{Gray}}c}

\usepackage{array}
\newcolumntype{L}[1]{$$>$${\raggedright\let\newline\\\arraybackslash\hspace{0pt}}m{#1}}
\newcolumntype{C}[1]{$$>$${\centering\let\newline\\\arraybackslash\hspace{0pt}}m{#1}}
\newcolumntype{R}[1]{$$>$${\raggedleft\let\newline\\\arraybackslash\hspace{0pt}}m{#1}}

\usepackage[colorlinks=true , citecolor=black,urlcolor=blue]{hyperref}

\newcommand{\xa}{\alpha}

\newcommand{\vxe}{\varepsilon}
\newcommand{\tm}{{\text -}}

\newcommand{\CC}{\R{CC}}

\newcommand{\tacq}{t_{\R{acq}}}
\newcommand{\xg}{\gamma}
\newcommand{\xt}{\vartheta}

\newcommand{\xk}{\kappa}

\newcommand{\xr}{\rho}

\newcommand{\nb}{\nabla}

\newcommand{\app}{\approx}

\newcommand{\Bp}{B_{\R{pol}}}

\newcommand{\Cs}{{}^{13}\R{C}}

\newcommand{\xy}[0]{\xhat\tm\yhat}

\newcommand{\mP}[0]{\mathcal{P}}

\newcommand{\fr}[2]{\frac{#1}{#2}}

\newcommand{\sq}[1]{\sqrt{#1}}

\newcommand{\mH}[0]{\mathcal{H}}

\newcommand{\rt}{\rightarrow}

\newcommand{\beq}{\begin{equation}}
\newcommand{\eeq}{\end{equation}}
                  
\newcommand{\benum}{\begin{enumerate}}
\newcommand{\eenum}{\end{enumerate}}
                    
\newcommand{\bit}{\begin{itemize}}
\newcommand{\eit}{\end{itemize}}
\newcommand{\xhat}{\hat{\T{x}}}
\newcommand{\yhat}{\hat{\T{y}}}

\newcommand{\bea}{\begin{eqnarray}}
\newcommand{\eea}{\end{eqnarray}}

\newcommand{\bev}{\begin{verbatim}}
\newcommand{\eev}{\end{verbatim}}

\newcommand{\qt}{\tau}

\newcommand{\lb}{\left(}
\newcommand{\rb}{\right)}
\newcommand{\lsb}{\left[}
\newcommand{\rsb}{\right]}

\newcommand{\T}[1]{\textbf{#1}}
\newcommand{\I}[1]{\textit{#1}}
\newcommand{\R}[1]{\textrm{#1}}

\newcommand{\zl}[1]{\label{eqn:#1}}
\newcommand{\zr}[1]{Eq.\,(\ref{eqn:#1})}
\newcommand{\zfl}[1]{\protect\label{fig:#1}}
\newcommand{\zfr}[1]{\figurename\,\ref{fig:#1}}

\newcommand{\ba}{\left\{ \begin{array}{lr}}
\newcommand{\ea}{\end{array}\right.}

\newcommand{\BRd}[1]{\textcolor{red}{#1}} 

\newcommand{\del}{\partial}

\newcommand{\blist}[1]{
 \begin{list}{#1}
 \begin{align}
	 arrow
 \end{align}
 $\checkmark\star
  { \setlength{\itemsep}{3pt}
     \setlength{\parsep}{2pt}
     \setlength{\topsep}{3pt}
     \setlength{\partopsep}{0pt}
     \setlength{\leftmargin}{1em}
     \setlength{\labelwidth}{1em}
     \setlength{\labelsep}{0.5em} } }
\newcommand{\elist}{
  \end{list}  }

\DeclareMathSymbol{\vartheta}{\mathalpha}{letters}{"12}
\DeclareMathSymbol{\theta}{\mathalpha}{letters}{"23}
\DeclareMathSymbol{\phi}{\mathalpha}{letters}{"27}
\DeclareMathSymbol{\varphi}{\mathalpha}{letters}{"1E}

\newcommand{\bef}
{
\begin{figure}[htbp]
\centering
}

\newcommand{\eef}{\end{figure}}

\addto\captionsenglish{\renewcommand{\figurename}{Fig.}}
\mleftright
\newcommand{\beginsupplement}{%
        \setcounter{table}{0}
        \renewcommand{\thetable}{S\arabic{table}}%
        \setcounter{figure}{0}
        \renewcommand{\thefigure}{S\arabic{figure}}%
				
     }

\newcommand{\affA}{Department of Chemistry, University of California, Berkeley, Berkeley, CA 94720, USA.}
\newcommand{\affB}{Chemical Sciences Division,  Lawrence Berkeley National Laboratory,  Berkeley, CA 94720, USA.}
\newcommand{\affC}{Fakultät Physik, Technische Universität Dortmund, D-44221 Dortmund, Germany.}

\begin{document}
\title{\T{Electron induced nanoscale nuclear spin relaxation probed by hyperpolarization injection}}
	
	\author{William Beatrez}\affiliation{\affA}
	\author{Arjun Pillai}\affiliation{\affA}
\author{Otto Janes}\affiliation{\affA}
\author{Dieter Suter}\affiliation{\affC}
\author{Ashok Ajoy}\email{ashokaj@berkeley.edu}\affiliation{\affA}\affiliation{\affB}

\begin{abstract}
We report on experiments that quantify the role of a central electronic spin as a relaxation source for nuclear spins in its nanoscale environment. Our strategy exploits hyperpolarization injection from the electron as a means to controllably probe an increasing number of nuclear spins in the bath, and subsequently interrogate them with high fidelity. Our experiments are focused on a model system of a nitrogen vacancy (NV) center electronic spin surrounded by several hundred $\Cs$ nuclear spins. We observe that the $\Cs$ transverse spin relaxation times vary significantly with the extent of hyperpolarization injection, allowing the ability to measure the influence of electron mediated relaxation extending over several nanometers.  These results suggest interesting new means to spatially discriminate nuclear spins in a nanoscale environment, and have direct relevance to dynamic nuclear polarization and quantum sensors and memories constructed from hyperpolarized nuclei.
\end{abstract}

\maketitle

\T{\I{Introduction}} --   Coupled electron-nuclear spin systems are highly  relevant to quantum information processing~\cite{prokofev00,Ladd2010}, quantum sensing~\cite{Degen17}, and dynamic nuclear polarization (DNP)~\cite{Abragam78}. Consider an electronic spin centrally located (position $r{=}0$) in a bath of nuclear spins within a magnetic field $B_0$ (see \zfr{fig1}A-B), coupled to them via hyperfine interactions. The electrons are fast-relaxing (short $T_{1e}$), often on account of interactions with phonons~\cite{Jarmola12}, and in turn they can serve as a relaxation source for the nuclear spins~\cite{Witzel06,bloembergen47,Blumberg60} (schematically shown in \zfr{fig1}C). Such electron induced nuclear relaxation is an important consideration for applications in quantum registers, memories, and sensors constructed out of nuclear spins~\cite{Hanson08,Reiserer16,Morton08}, and is therefore important to quantify. It also plays a key role in determining rates of polarization transfer in DNP~\cite{Siaw14, Saliba17,Sarkar21}. However, probing such relaxation influences, particularly in a spatially defined manner, is challenging. This is because experiments typically have very restricted access to spins in the bath, \I{i.e.} there is a limited possibility of spatially distinguishing the spins, other than a small shell where the nuclear resonance frequencies can be significantly shifted from those of the bulk~\cite{Blumberg60,Tse68,Horvitz71,Stern21,Ramanathan08}. Previous nanoscale quantum sensing experiments have, for instance, been limited to proximal central spin relaxation effects in small ($<$20) spin networks~\cite{Dreau14,Abobeih19}.

In this paper, we report on experiments that study the effects of an electronic spin on nuclei over wider length scales, spanning several nanometers and involving several hundred nuclei. Our strategy (\zfr{fig1}D) exploits controllable hyperpolarized spin injection from the electron to the nuclear bath. The hyperpolarization time $\tau$ is employed as a knob to tune length scales in the bath being probed; the polarization is carried over longer distances with increasing $\tau$. Simultaneously, the nuclei are subject to a RF driving protocol that permits continuous tracking of their magnetization over minute long periods with high signal-to-noise (SNR)~\cite{Sahin22}. The measurements reveal, surprisingly, that ensemble-averaged nuclear lifetimes $T_2^\prime$ in the rotating frame increase dramatically with increasing polarization time $\tau$. We demonstrate that this constitutes a direct experimental signature of electron induced nuclear spin relaxation, allowing us to quantify its influence extending over several nanometers.   

\begin{figure}[t]
  \centering
  {\includegraphics[width = 0.485\textwidth]{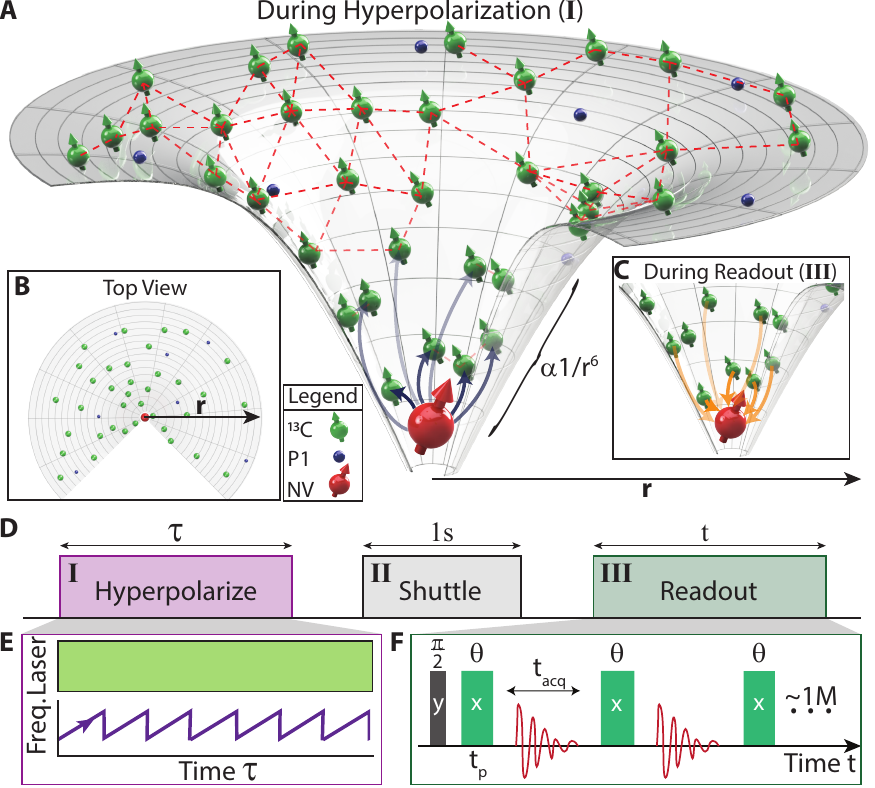}}
  \caption{\T{System and protocol. } (A-C) \I{System} consists of central NV electron (red), $\Cs$ nuclei (green), and P1 centers (blue) at distance $r$. Dashed lines are inter-nuclear dipolar couplings. Profile denotes $1/r^6$ NV-mediated relaxation effect. (A,C) Role of NV center as polarization source and relaxation sink respectively, during different experimental regimes. (D) \I{Experiment schematic:}\T{(I)} NV${\rt}\Cs$ hyperpolarization for period $\tau$ at 36mT, \T{(II)} transport to high field and \T{(III)} $\Cs$ readout for time $t$ at 7T. (E) \I{Hyperpolarization} (\T{I}) involves MW chirps~\cite{Ajoy17, Ajoy18}. (F) \I{Measurement} (\T{III}) comprises a train of spin-locking $\xt$-pulses with interrogation in $\tacq$ inter-pulse intervals~\cite{Beatrez21}. \I{Note}: Throughout the manuscript, time $\tau$ describes the hyperpolarization (spin-lock) time and time $t$ describes the readout (detection) time. }
\zfl{fig1}
\end{figure}

\begin{figure}[t]
  \centering
  {\includegraphics[width = 0.485\textwidth]{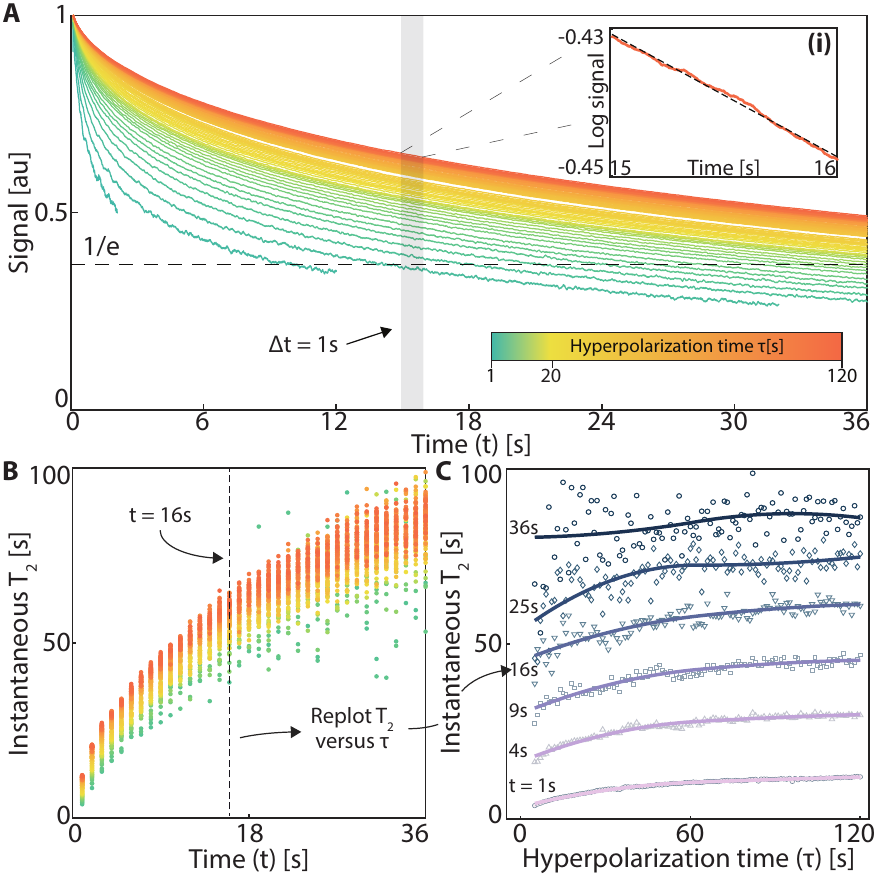}}
  \caption{\T{Effect of increasing hyperpolarization time $\tau$. } (A) \I{Spin-lock decays} for different hyperpolarization times $\qt$ corresponding to colorbar (regime \T{I} in \zfr{fig1}D). Single-shot data here (obtained with $\tacq{=}32\mu$s and $\xt{\app}\pi/2$ in \zfr{fig1}F) is boxcar averaged over 97ms, and normalized after truncation at $t{=}9.7$ms. Data for $t{=}36$s corresponds to $\sim$369k pulses (see \zfr{fig3} for full data). Dashed line represents $1/e$ intercept. Measured $T_2^\prime$ increases ${\app}6.42$-fold between $\qt{=}2$s and $\qt{=}120s$.  \I{Inset (i):} Zoom into representative 1s segment (gray window in (A)). Dashed line is a fitted piecewise monoexponential. (B) \I{Instantaneous lifetime $T_2$} measured along decay curve, extracted from slopes as in A(i). Color bar (see (A)) represents $\qt$. Dashed line shows exemplary segment ending at 16s (gray window in (A)). (C) \I{Instantaneous $T_2$} plotted against $\qt$. Points show $T_2$ lifetimes from (B) for exemplary 1s segments ending at labeled $t$ values (different markers). Lines are spline fits.}
\zfl{fig2}
\end{figure}

\begin{figure*}[t]
  \centering
  {\includegraphics[width = 1\textwidth]{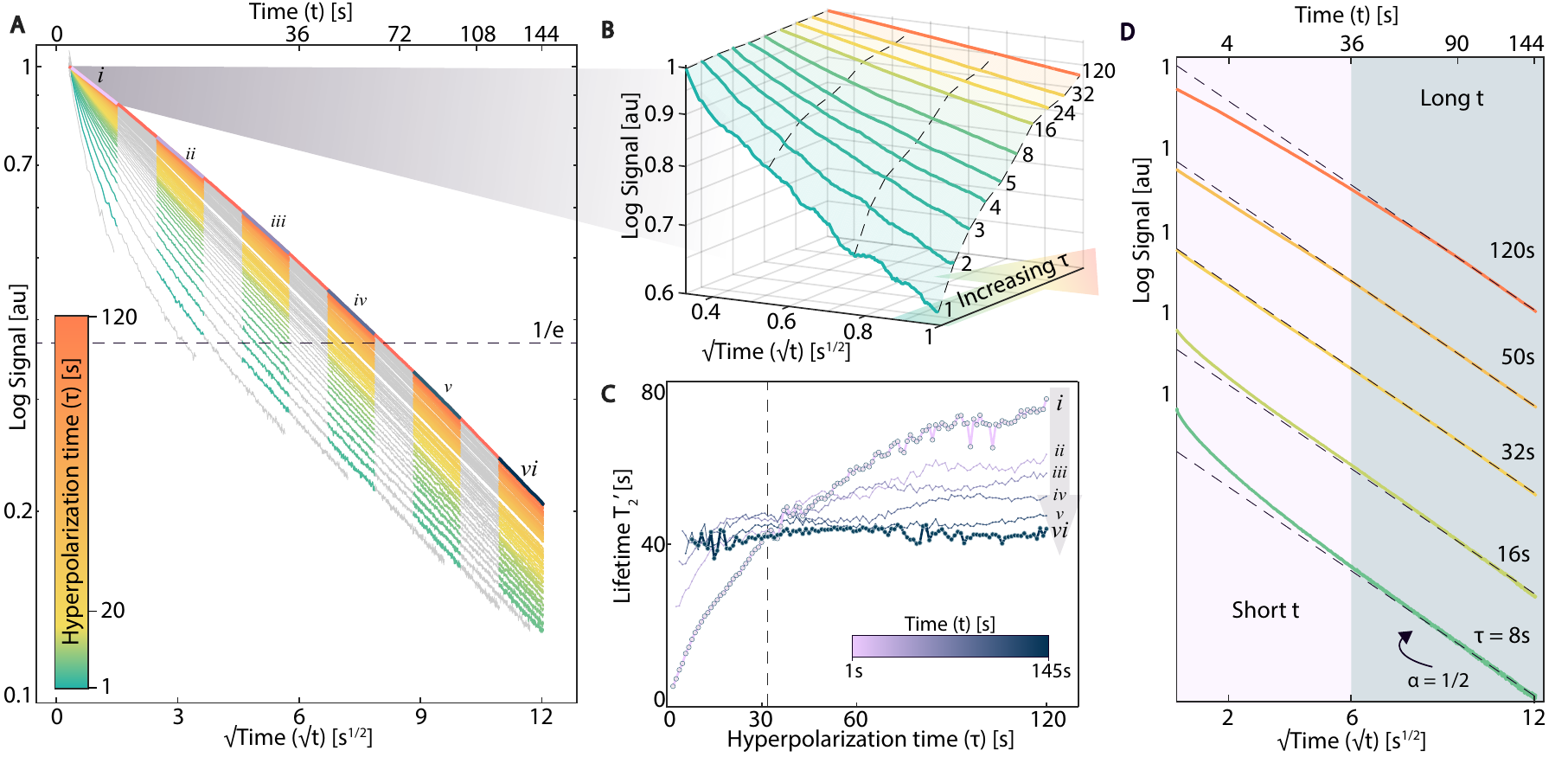}}
  \caption{\T{Quantifying instantaneous decay rates. } (A) \I{Full data} in \zfr{fig2} plotted on a logarithmic scale with respect to $\sqrt t$ for different $\qt$ (color bar). Traces extending to $t{=}144$s have ${\sim} 10^6$ points. Upper axis represents time $t$. Six equally sized (in $\sqrt t$) segments (labeled \I{(i)}-\I{(vi)}) are selected along the line. (B) \I{Zoom into segment \I{(i)}} (sampled every $\app$5ms) for representative marked $\tau$ values (same colorbar as in (A)). Dashed lines are at equally spaced $\sq{t}$ values to guide the eye. (C) \I{Extracted time constants $T_2^\prime$} from slopes of the corresponding segments in (A). Segment \I{(i)} shows a steep increase in $T_2^\prime$ with $\qt$; subsequent segments show progressively flatter profiles (gray arrow). Crossover of early and late segments occurs at $\tau{\app}32$s (dashed line). (D) \I{Variations in short and long time decay} behavior with $\tau$. At long $t$ (dark shaded region), decay closely follows $\alpha {=} 1/2$ (dashed lines, extracted from segment \I{(vi)}), while short times $t$ show a transition from convex to concave behavior. Transition occurs at $\tau{\app}32s$ (similar to \zfr{fig3}C). Upper axis represents time $t$. Line colors correspond to colorbar in panel (A). }
\zfl{fig3}
\end{figure*}

\begin{figure*}[t]
  \centering
  {\includegraphics[width = 1\textwidth]{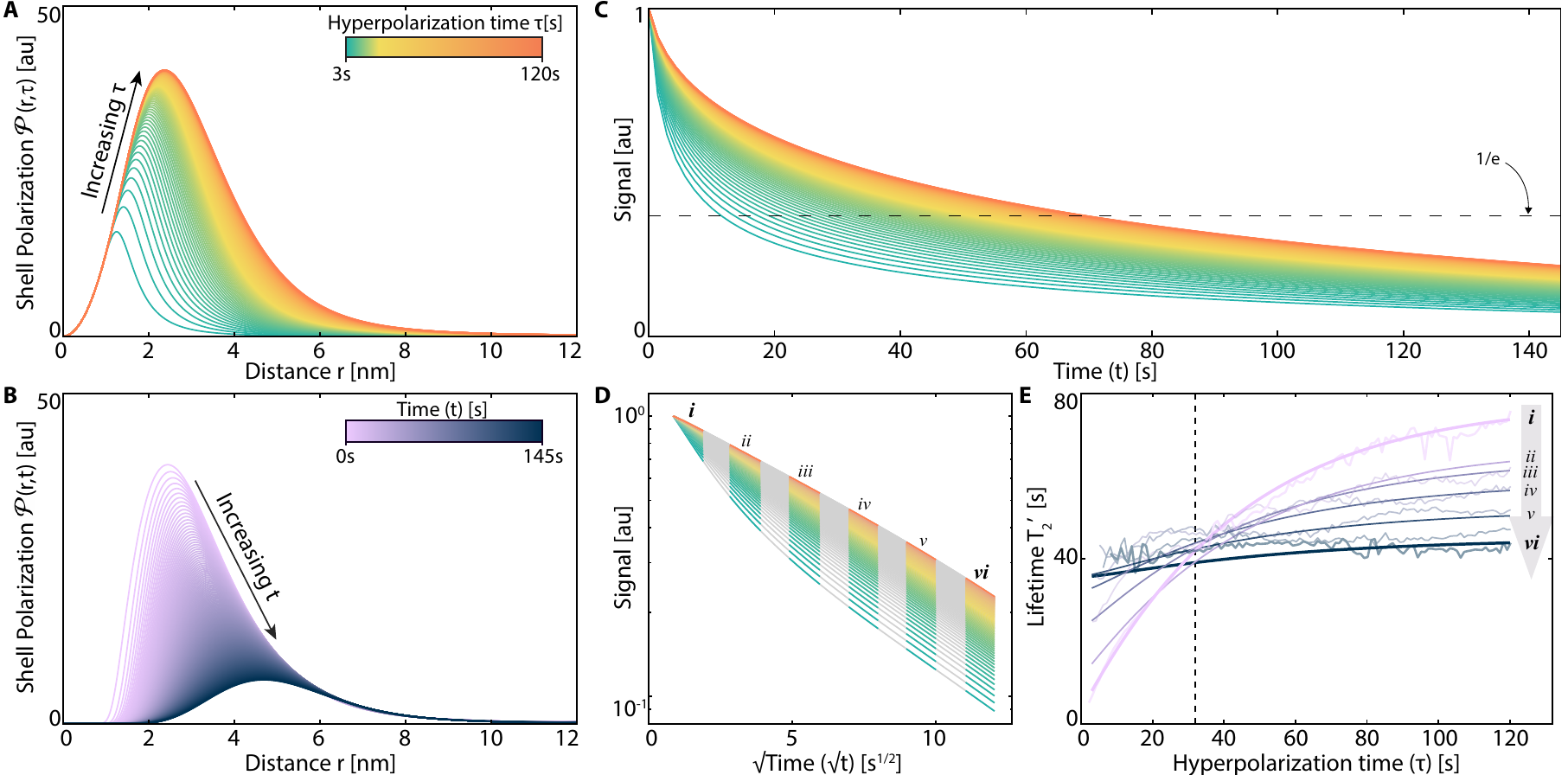}}
  \caption{\T{ Simulations of polarization evolution in time and space}.  (A) \I{Spatial polarization build-up} with $\qt$ (colorbar) plotting polarization $\mP(r,\qt)$ in a shell at distance $r$ from the central NV center (during regime \T{I} in \zfr{fig1}D).
	(B) \I{Polarization distribution during readout} period $t$ (colorbar) in regime \T{III} of \zfr{fig1}D. Here, we set the initial condition to be the $\tau {=} 120$s polarization distribution in panel (A). 
	(C) \I{Simulated NMR signals} following panels (A-B) for different hyperpolarization times $\qt$ similar to \zfr{fig2}A. Data reveals $T_2^\prime$ increase with $\tau$, in agreement with the experimental data in \zfr{fig2} (see (A) for colorbar).
	(D) \I{Stretched exponential decay dynamics}. Analogous to \zfr{fig3}A, data from (C) is plotted on a logarithmic scale with respect to $\sqrt{t}$.  Stretched exponential behavior qualitatively matches data in \zfr{fig3}A. Segments labeled \I{(i)-(vi)} are used to extract $T_2^\prime$ decay constants.
	(E) \I{Extracted $T_2^\prime$ decay constants} for representative segments \I{(i)}-\I{(vi)} show qualitative agreement with \zfr{fig3}C. Here, dark lines are simulation and light lines are experimental data. Dashed line marks crossover at $\qt{\sim}32$s.}
\zfl{fig4}
\end{figure*}

\T{\I{System and Protocol}} -- Experiments here are carried out in diamond as a model system, with central NV~\cite{Jelezko06,Manson06} electronic spins surrounded by $\Cs$ nuclei (\zfr{fig1}A-B).  At ${\app}$1ppm, the NV centers are separated by ${\app}12$nm, and $\Cs$ nuclei appear with lattice density ${\app}$0.92 spins/nm$^3$, yielding a relative NV:$\Cs$ ratio ${\sim}1{:}10^4$~\cite{Ajoy19relax,Wyk97}. In addition, the lattice hosts randomly positioned paramagnetic impurities (P1 centers) at a concentration $>$20ppm, which can also serve as relaxation sources (\zfr{fig1}A-B). 

Experiments are conducted in three regimes (indicated in \zfr{fig1}D): \T{(I)} optically induced NV${\rt}\Cs$ spin injection for period $\tau$ at low-field ($\Bp{=}36$mT), \T{(II)} rapid adiabatic transport to high field, and \T{(III)} $\Cs$ interrogation at $B_0{=}7$T.  For hyperpolarization (\zfr{fig1}E), we exploit a mechanism previously described in Ref.~\cite{Ajoy17,Ajoy18,Ajoy20cube}. Distant spins are polarized by spin diffusion driven by the internuclear dipolar Hamiltonian~\cite{Ramanathan08,Arabanov15,Hovav10}, $\mH_{\R{dd}} {=} \sum_{j<k} d_{jk}^{\CC}(3I_{jz}I_{kz} - \vec{I_j}\cdot\vec{I_k})$, where $I_j$ refer to $j^{\R{th}}$ spin-1/2 operators, and coupling strengths $d_{jk}^{\CC}{\propto} \xg_n^2/r^3$, with magnetogyric ratio $\xg_n{=}$10.7MHz/T, and $r$ being the inter-nuclear distance. Bulk averaged $\Cs$ hyperpolarization under maximal conditions is $\vxe{\app}0.3\%$.

Sample transport (regime \T{II}) occurs in $t_s{\app}1\R{s}{\ll} T_{1n}$ in a manner adiabatic with respect to the instantaneous $\Cs$ Larmor frequency, thereby preserving the hyperpolarization generated~\cite{Ajoyinstrument18}. Conversely, since $t_s{\gg} T_{1e}{\sim}1$ms~\cite{Popa04,Jarmola12}, the NV center rapidly loses hyperpolarization and ultimately returns to thermal polarization levels ${\app}3$\% in regime \T{III} (\zfr{fig1}F). Subsequently, the NV center predominantly serves as a point relaxation source for the nuclear bath (\zfr{fig1}C).

Ensemble $\Cs$ readout (regime \T{III}) employs a protocol described in Refs. \cite{AjoyDD20, Beatrez21}. $\Cs$ spins are prepared along the transverse axis $\hat{\T{x}}$ ($\xr_I{=}\vxe \sum_{j}I_{jx}$) on the Bloch sphere, and a train of spin-locking $\xt$-pulses are applied~\cite{Rhim76,Rhim73b}. $\Cs$ nuclei are interrogated in windows between the pulses, allowing their dynamics to be continuously tracked with high SNR~\cite{SOM}.  Signal obtained corresponds to measuring the survival probability in the $\xy$ plane. The sequence operation can be described by the unitary $U(t)$. For sufficiently rapid pulsing duty cycle,  $U(t){\app}\exp(i{\mH}^{(0)}_{F}t)$, such that the inter-nuclear Hamiltonian is engineered to leading order in the Magnus expansion to 
${\mH}^{(0)}_{F} {\app} \sum_{j<k}d_{jk}^{\CC}\lb \fr{3}{2}\mH_{\R{ff}}-\vec{I_j}\cdot\vec{I_k}\rb$, 
where $\mH_{\R{ff}}{=} I_{jz}I_{kz}+ I_{jy}I_{ky}$ is a flip-flop Hamiltonian ~\cite{Beatrez21}. Since $[\xr_I{,}{\mH}^{(0)}_{F}]{=}0$ commutes with the initial state, dipolar evolution is suppressed to leading order. As a result, $\Cs$ free induction decay lifetimes $T_2^*{\sim} 1.5$ms (in the absence of spin-lock pulses), are significantly extended, here to $T_2^\prime{\gtrsim}$65.5s (see \zfr{fig2}A).

\T{\I{Results}} —  \zfr{fig2}A describes our primary experimental result (see movie at Ref. ~\cite{SD_video1}), showing $\Cs$ NMR signal measured employing differing hyperpolarization periods $\tau$ (varied every one second from $\tau{=}$1s to $\tau{=}120$s, see colorbar). The signals here are normalized to their values at 9.7ms. Each 36s trace consists of ${\sim}369,000$ pulses (full data in \zfr{fig3} consists of ${\sim}{10^{6}}$ pulses), and the $\Cs$ nuclei are interrogated after every pulse (\zfr{fig1}F). Surprisingly, we observe that the signals decay \I{more slowly} with increasing hyperpolarization period $\tau$. Normalization of the signal profiles allows the ability to unravel the \I{relative} changes in the decay time constant $T_2^\prime$, estimated from the $1/e$ intercept (dashed line in \zfr{fig2}A). For example, comparing $\tau{=}2$s and $\tau{=}$120s in \zfr{fig2}A, we observe a $T_2^\prime$ increase from 10.2s to 65.5s.

The $1/e$-intercept (dashed line in \zfr{fig2}A) is blind to the instantaneous change of the decay profile and therefore provides only limited information. To more clearly observe the decay dynamics, we divide the curves in \zfr{fig2}A into 36 segments of width $\Delta{t}{=}1$s, with one such segment shown in the gray window in \zfr{fig2}A. We fit the decay in each segment to a monoexponential (as in inset (i)) and extract the \I{instantaneous} time constants $T_2(t)$, which are plotted in \zfr{fig2}B. The notation $T_2$ (as opposed to $T_2^\prime$) is used to emphasize that these are monoexponential constants.  Colors here represent $\qt$ with the same color bar as in \zfr{fig2}A.  For each trace, the signals decay markedly slower with increasing time $t$ (see also \zfr{fig3}).  Ultimately,  the $T_2$ times are remarkably long ($T_2{\app}$100s) at large ${\tau}$. Increasing polarization time $\qt$ makes the overall decay slower for any selected segment. Indeed, for the segment ending at $t{=}16$s (vertical dashed line in \zfr{fig2}B), the $T_2$ value is increased by $1.66$-fold. To now emphasize the relative change in the $T_2$ values for different segments, \zfr{fig2}C shows the $T_2$ lifetimes plotted against $\tau$, where data corresponding to each segment in \zfr{fig2}A-B forms the lines (values denote segments ending at $t$). \zfr{fig2}C makes clear that the instantaneous $T_2$ increases for each segment with increasing $t$, and within each segment with hyperpolarization time $\tau$. The maximum relative change occurs for short $\qt$ and at early $t$. 

A clearer view of data in \zfr{fig2} can be obtained by noting that the decays approximately follow a stretched exponential of the form $\exp\lsb -(t/T_2^\prime)^{\alpha}\rsb$ with $\alpha{\app}1/2$ (see \zfr{fig3} and movies at Refs. ~\cite{SD_video2, SD_video3}). \zfr{fig3}A shows the $\Cs$ signals in \zfr{fig2} plotted on a logarithmic scale with respect to $\sqrt{t}$; the colorbar represents increasing hyperpolarization time $\tau$. The signals (gray lines) then manifest as approximately straight lines, demonstrating stretched exponential character with $\alpha{\app}1/2$. $T_2^\prime$ lifetimes can now be extracted conveniently from the instantaneous slopes $s$, as $T_2^\prime{=}1/s^2$. High SNR, along with rapid data collection rates (allowing ${\sim}10^6$ points per trace), allow the unique ability to continuously observe the stretched exponential dynamics. We now focus attention to six segments along the decay curves (labeled \I{(i)}-\I{(vi)} in \zfr{fig3}A). Complementary to \zfr{fig2}A, data reveals that the $T_2^\prime$ values at short time $t$ vary considerably with increasing $\tau$, as evident in segment \I{(i)}. In contrast, the $T_2^\prime$ values at long time $t$ are observed to be independent of $\tau$ (evident from the approximately parallel relaxation profiles in segment \I{(vi)}). Segment \I{(i)} is zoomed in \zfr{fig3}B on the same logarithmic scale for representative values of $\qt$. The panel makes clear the large relative change in $T_2^\prime$ with $\tau$,  most evident in the difference between traces corresponding to $\tau {=}$1s and $\qt{=}$120s.

To quantify this change with greater clarity, \zfr{fig3}C elucidates the extracted $T_2^\prime$ values for the six segments, with a special emphasis on segments \I{(i)} (0-1s) and \I{(vi)} (121-144s) (highlighted traces). Data reveals that the $T_2^{\prime(i)}$ grows significantly with increasing hyperpolarization time $\qt$, while $T_2^{\prime(vi)}$ is flat and almost independent of $\tau$. Indeed, the $\tau$ dependence of $T_2^\prime$ becomes \I{“flatter”} with increasing segment number. Interestingly also, the extracted $T_2$ values for the different segments display a crossing point at $\tau{\app}$32s (dashed line in \zfr{fig3}C). Finally, \zfr{fig3}D offers insight into variations from exact $\alpha{=}1/2$ stretched exponential behavior with changing $\tau$.  We observe that at long times $t$ (dark shaded region), the behavior follows a \I{universal} $\xa{=}1/2$ dependence (dashed lines). However, in the short time region ($t{\lesssim}$36s), there is a deviation from this, exhibited as a transition from convex to concave behavior around $\alpha{=}1/2$. Cross-over occurs at $\tau{\app}$32s, similar to \zfr{fig3}C. 

While similar stretched exponential decay dynamics have been observed before in some restrictive model systems~\cite{Lock92,Hartman94,Henrichs84}, the relatively low number of measurement points in these reports allowed only a limited ability to characterize them. By contrast, the ${\sim}10^6$ points per trace in \zfr{fig3}A, along with the high SNR, provide a remarkably clear view into the dynamics at short and long $t$.

\T{\I{Theory}} --
 To describe the experimental observations, we construct a simple, semi-quantitative model for the nuclear polarization $p(r,t)$ at coordinate $r$ and time $t$. We assume centrosymmetry, a good approximation given the ensemble average in our experiments. We then model the dynamics using the differential equation,
\beq
\fr{\del}{\del t}p(r,t) = \fr{P_0}{r^6} -\fr{\kappa_0}{r^6}p(r,t) -\fr{1}{T_1}p(r,t) +D\nb^2 p(r,t)
\zl{SD}
\eeq
where $P_0$ denotes the rate of hyperpolarization injection and $\xk_0$ is the strength of spatially dependent relaxation due to the central NV center.  The $1/r^6$ scaling of the $P_0$ term in \zr{SD} does not yield qualitatively different behavior compared to a $1/r^3$ scaling but makes the equation better conditioned near $r{=}0$. In any case, the $r{\rt}0$ spatial regime and associated nuclear spin frozen core are outside the ambit of our experimental observations because of the relatively small $\Cs$ detection bandwidth (${\app}$32kHz) in our measurements. In contrast to the NV center (at $r{=}0$), we assume that the relatively dense P1 centers (${>}$20ppm) serve as contributors to \I{background} relaxation of the $\Cs$ nuclei independent of their position; this is captured by the $T_1$ term in \zr{SD}. Finally, the last term denotes spin diffusion which we assume to be Fickian and Gaussian with a single constant $D$ at all locations. This is a good approximation given the large number of $\Cs$ nuclei participating in the measured signal around every NV center~\cite{Ramanathan08}. For instance, a sphere of injected polarization with radius 4nm contains ${\app}247$ nuclear spins. 

We solve \zr{SD} separately in the three regimes of the experiment (\zfr{fig1}D), with the solution for one regime setting the initial conditions for the next. Obtaining a precise estimate of the parameters in \zr{SD} is challenging and outside the scope of this paper; our goals instead are to obtain \I{qualitative} agreement with the experimental observations in \zfr{fig2}-\zfr{fig3}. We therefore make some simplifying assumptions to the parameters in \zr{SD}. In regime \T{I}, we assume $P_0/\xk_0{=}1/100$ for simplicity ($P_0{=}0$ in regime \T{III}). In regime \T{I} and \T{III}, we employ $T_1{=}q^{-1}T_{1,\R{LF}}$ and $T_1{=}q^{-1}T_{1,\R{HF}}$ respectively, where $T_{1,\R{LF}}{=}283$s and $T_{1,\R{HF}}{=}1520$s are the measured low and high-field bulk $\Cs$ lifetimes respectively at $\qt{=}60$s~\cite{SOM}, and $q$ is a scaling factor that we employ as a free parameter in the fits. For \zfr{fig4}, we find good agreement with $q{=}6.75$. The latter assumption can be rationalized by the fact that (1) the individual contributions from P1 centers to $T_1$ relaxation are hard to separate in the bulk $T_1$ measurements; (2) longitudinal and transverse relaxation rates in regime \T{III} are expected to be proportional, but are in general, unequal, making the measured $T_{1,\R{HF}}$ only approximately reflective of $T_1$ in \zr{SD}; and (3) $T_{1,\R{LF}}$ is measured under dark conditions at a fixed $\qt{=}60$s, but the corresponding $T_1$ in regime \T{I} is measured under optical illumination (making measurement more challenging). 

With these assumptions, \zfr{fig4}A-B shows the simulated system dynamics, where we plot the polarization contained in a shell at radius $r$, $\mP{=}4\pi r^2 p$. \zfr{fig4}A first shows $\mP(r,\qt)$ with increasing hyperpolarization time $\qt$ (see colorbar) in regime $\T{I}$, assuming a 1s shuttling period in regime $\T{II}$. Simulation parameters are set to obtain good qualitative agreement with experiment (here $D{=}0.0135$nm$^{2}$/s and $P_0{=}6.75$s$^{-1}$). With increasing $\qt$, spin diffusion leads to a spread of polarization; this is evidenced by the movement of the "centroid" of $\mP(r,\tau)$ with increasing $\qt$ towards greater $r$ in \zfr{fig4}A. Notably however, the continuous replenishment of polarization from the NV in regime \T{I} makes the distribution of $\mP(r,\qt)$ skewed towards the left. In a complementary manner, \zfr{fig4}B elucidates the distribution $\mP(r,t)$ during the readout period in regime \T{III}, assuming one starts with the distribution obtained with $\qt{=}120$s (right-most trace in \zfr{fig4}A). The strong relaxing effect of the NV center yields the polarization “hole” close to $r{=}0$, and manifests as the steep wall of growing polarization in \zfr{fig4}B. Additionally, the centroid of $\mP(r,t)$ moves towards larger $r$ and homogenizes as $t$ increases; the shift with $r$ is apparently larger here because there is no polarization replenishment and relaxation is slower for larger $r$. 

\zfr{fig4}C displays the net polarization $\int\mP(r,t)dr$ following the trajectory of \zfr{fig4}B as interrogated during the readout period, but considering here the result for different hyperpolarization times $\qt$ (see colorbar in \zfr{fig4}A). Normalizing the traces then yields an apparent polarization relaxation that becomes slower with increasing $\qt$, closely matching the experimental results in \zfr{fig2}A. From \zfr{fig4}A-B, we identify that the observations in \zfr{fig2}A arise because the shifting centroid of $\mP$ (see \zfr{fig4}A) makes electron mediated relaxation less effective with increasing $\qt$. Interestingly, we find that the decays in \zfr{fig4}A also follow an identical $\xa{=}1/2$ stretched exponential as in the experiment. This is evidenced by replotting the data in a logarithmic scale with respect to $\sq{t}$ in \zfr{fig4}D, where we observe a behavior similar to the experiments in \zfr{fig3}A: the stretched exponential dynamics manifest as straight lines here. We hypothesize that the stretched exponentials here result from the inhomogeneity of the polarization distribution in \zfr{fig4}A, which is subject to a collection of relaxation rates based on spatial proximity to the NV. At long readout periods $t$, the parallel lines yield a universal relaxation profile that is independent of $\qt$; \zfr{fig4}B allows us to recognize that this is because the polarization spreading far from the NV center is predominantly influenced by the background $1/T_1$ relaxation. Indeed, as shown in \zfr{fig4}E, taking segments $\I{(i)-(vi)}$ along the decay curves in \zfr{fig4}D, and extracting their relaxation times $T_2^\prime$, we observe a progressive flattening of the $T_2^\prime$ values with increasing $\qt$, in good agreement with experimental data in \zfr{fig3}C (overlapped here). Cross-over occurs at $\qt{\sim}32$s similar to \zfr{fig3}C.

\T{\I{Outlook}} -- These results open many interesting future avenues in a wide class of central spin systems. First, as \zfr{fig3}A and \zfr{fig4} demonstrate, spin-lock control allows for peering into nuclear polarization localized at different positions $r$ with respect to the central electron. Moving through time $t$, the decay curve, as seen in \zfr{fig2}A, effectively corresponds to shifting the centroid location of the sensitive region being probed in the lattice (\zfr{fig4}B). This suggests a (nonlinear) means to map from $t$ to an effective $r$ coordinate, and can allow interesting new ways to discriminate nuclear spins in the electronic environment. It can enable visualization of the transport of polarization from the central electron~\cite{Zhang98,Burgarth17,Ajoy12,Boutis04}, and the ability to rewritably engineer spin texture in the nuclear bath~\cite{King12,Michal99}. For instance, successive injection of positive and negative polarization from the NV center can  result in non-equilibrium “shells” of nuclear polarization whose dynamics can be frozen for minute-long periods by decoupling. Ultimately, this engenders opportunities for exploiting mesoscale nuclear spin baths for quantum information science (QIS)~\cite{Foletti09} and sensing~\cite{Sahin21,Ajoy12gyro,Jaskula19,Ledbetter2012}.

Our results illustrate the ability of nuclear spins to probe electronic relaxation processes, therefore providing a means to view the phonon density of states that dominate these relaxation mechanisms~\cite{Jarmola12,Takahashi08}. This is relevant to the emerging field of QIS with molecular systems~\cite{Bayliss20}, where there is an important need to develop design rules for electronic lifetimes based on molecular vibronic properties~\cite{He20}. It also has important implications for DNP, wherein the concentration and identity of electron spins can vastly affect nuclear polarization levels~\cite{Lange12}. Concurrently, this suggests exciting potential for leveraging long nuclear $T_2'$ lifetimes in systems capable of “turning off” the electron spin after hyperpolarization injection~\cite{Capozzi17}, for instance in molecules hosting triplet photoexcitable electrons~\cite{Henstra14,Tateishi14,Niketic15}.

\T{\I{Conclusion}} -- In summary, we have experimentally measured the influence of a central electronic spin in relaxing the surrounding nuclear spin bath. We exploited controlled hyperpolarization injection as a knob to control length scales over which bath spins are probed, extending over several nanometers and hundreds of spins. Our work informs on a new means to probe nanoscale spin environments, and portends applications in quantum memories and sensors constructed out of hyperpolarized nuclei. 

We gratefully acknowledge discussions with C. Ramanathan, C. Meriles and J. Reimer. This work was funded by ONR (N00014-20-1-2806) and DOE STTR (DE-SC0022441).

\bibliography{main_PRL.bbl}

\pagebreak

\clearpage
\onecolumngrid
\begin{center}
\textbf{\large{\textit{Supplementary Information} \\ \smallskip
Electron induced nanoscale nuclear spin relaxation probed by hyperpolarization injection}}\\
\hfill \break
\smallskip
William Beatrez,$^{1}$  Arjun Pillai,$^{1}$  Otto Janes,$^{1}$Dieter Suter,$^{2}$ and Ashok Ajoy,$^{1,3,\BRd{\ast}}$\\
\smallskip
\emph{${}^{1}$ {\small Department of Chemistry, University of California, Berkeley, Berkeley, CA 94720, USA.}} \\
\emph{${}^{2}$ {\small Fakultät Physik, Technische Universität Dortmund, D-44221 Dortmund, Germany.}} \\
\emph{${}^{3}$ {\small Chemical Science Division, Lawrence Berkeley National Laboratory, University of California, Berkeley, Berkeley, CA 94720, USA.}}
\end{center}

\twocolumngrid

\beginsupplement

\begin{figure}[t]
  \centering
  {\includegraphics[width = 0.4\textwidth]{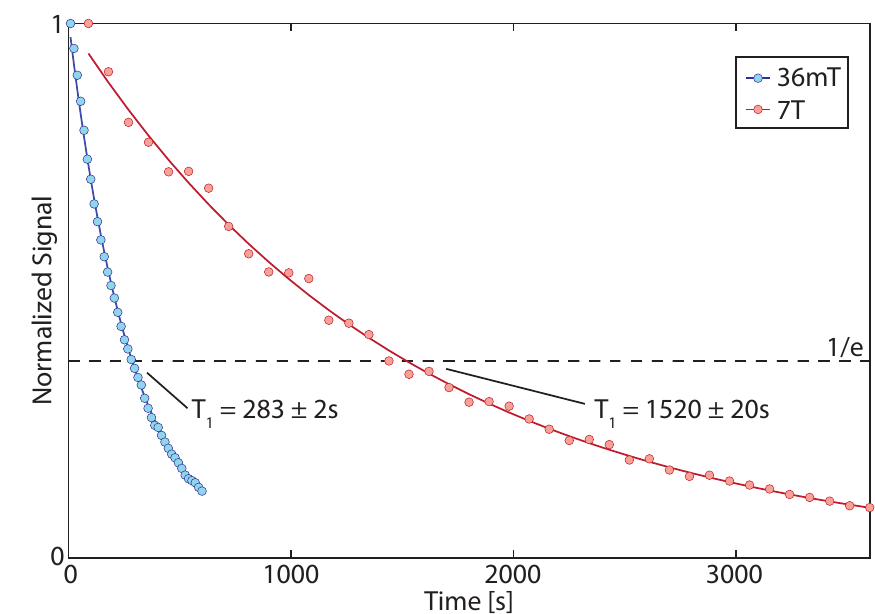}}
  \caption{\T{$T_1$ measurements. } \I{Sample $T_1$} at low-field (36mT, blue) and high-field (7T, red) at $\qt{=}60$s. Points are data and solid lines are monoexponential fits. Dashed line denotes $1/e$, and intersections give a low-field $T_1$ of 283$\pm$2s and a high field $T_1$ of 1520s$\pm$20s.  }
\zfl{figS1}
\end{figure}

\begin{figure}[t]
  \centering
  {\includegraphics[width = 0.5\textwidth]{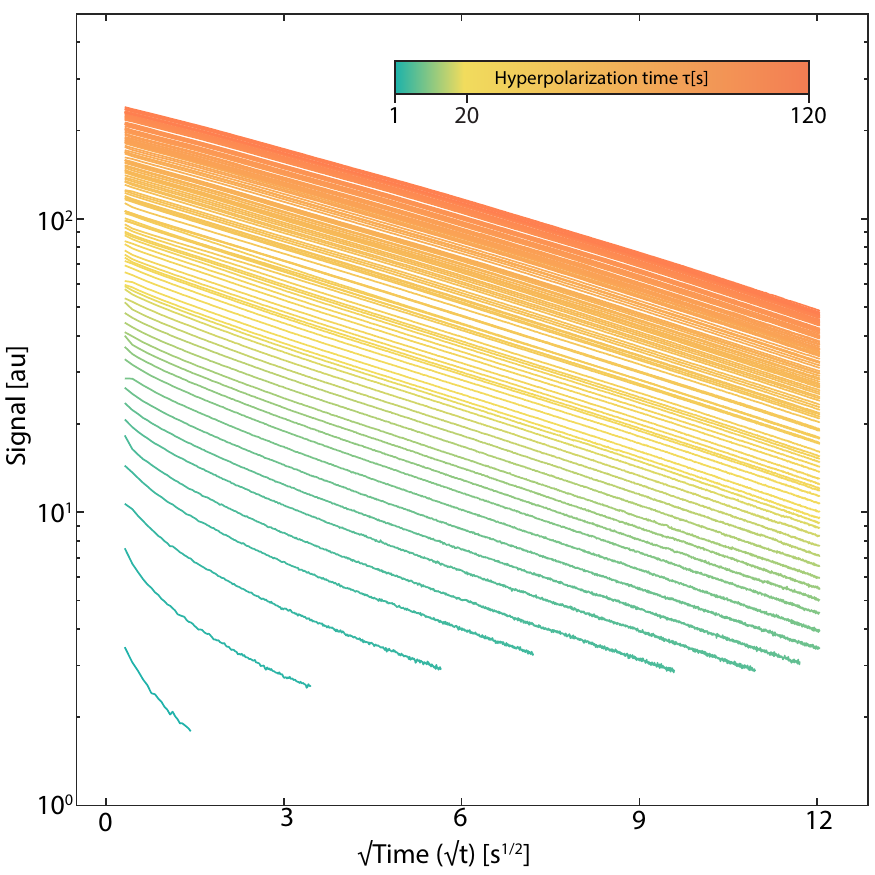}}
  \caption{\T{Polarization buildup with hyperpolarization period $\qt$}. Panel shows the unnormalized data corresponding to \zfr{fig3}A of the main paper. The different curves are the $\Cs$ pulsed spin-lock decays obtained under varying hyperpolarization time $\qt$ (see colorbar). This data is also shown as a movie in Ref.~\cite{SD_video3}.}
\zfl{figS3}
\end{figure}

\section{\texorpdfstring{$T_1$}{} measurements at low and high field}
\zfr{figS1} considers the measurement of the $T_1$ of the $\Cs$ nuclear spins at the polarizing field ($\Bp{=}36$mT) and the readout field ($B_0{=}7$T) respectively. For these measurements, the sample is first hyperpolarized, and is subsequently carried to the field of interest ($\Bp$ or $B_0$) for a waiting period $t_0$. Finally, the sample is transported to $B_0$ where spin-lock readout is carried out similar to \zfr{fig2}, and the integrated signal is plotted. The rapid field cycling for these measurements is carried out via mechanical sample shuttling, and the experimental strategy is similar to that in Ref.~\cite{Ajoy19relax}. 

Notably, varying the period $t_0$ allows one to discern the profile of the $T_1$ relaxation at both fields (see \zfr{figS1}). We find that the relaxation profiles closely follow monoexponential decay profiles (solid lines). From a $1/e$ intercept (dashed line) in \zfr{figS1} we find the respective low and high field relaxation times as $T_{1,\R{LF}}{=}283{\pm}$2s and $T_{1,\R{HF}}{=}1520\pm20$s respectively. We note that the monoexponential decays here stand in contrast to the observed stretched exponential decays that we find for the rotating frame lifetimes $T_2^\prime$ in \zfr{fig2} and \zfr{fig3} of the main paper.

\section{Movies showing data in \texorpdfstring{\zfr{fig2}}{} and \texorpdfstring{\zfr{fig3}}{}}
As a complement to the graphs in the main paper, we present movies corresponding to the data in \zfr{fig2}A and \zfr{fig3}A on Youtube (found at Refs.~\cite{SD_video1, SD_video2}). These movies show clearly the progressive slowing down of the decay profiles upon increasing the hyperpolarization time $\qt$ in both representations. The gray lines here show fitted stretched exponential lines corresponding to the previous data for clarity, allowing a guide to the eye to track the slowing relaxation dynamics with increasing $\qt$. \zfr{figS3} shows an unnormalized plot of \zfr{fig3}, from where the polarization buildup dynamics as a function of $\qt$ can be extracted.

\end{document}